\documentclass[conference]{IEEEtran}
\IEEEoverridecommandlockouts
\usepackage{placeins}
\usepackage{array} 
\usepackage{cite}
\usepackage{amsmath,amssymb,amsfonts}
\usepackage{algorithmic}
\usepackage{graphicx}
\usepackage{textcomp}
\usepackage{xcolor}
\definecolor{deepgreen}{rgb}{0,0.439,0.227}
\usepackage{enumitem}
\usepackage{amsmath, amssymb}
\usepackage{svg}
\usepackage{float}
\usepackage[linesnumbered,ruled,vlined]{algorithm2e}
\SetAlgoSkip{smallskip}

\usepackage{caption}
\usepackage{subcaption}
\usepackage{siunitx}
\usepackage{graphicx,tikz,adjustbox}
\usetikzlibrary{calc}
\RestyleAlgo{ruled}
\usepackage{multirow}
\usepackage{booktabs}
\usepackage{array}

\usepackage{xcolor}

\IEEEsettopmargin{t}{0.75in}
\SetKw{KwBy}{by}
\def\BibTeX{{\rm B\kern-.05em{\sc i\kern-.025em b}\kern-.08em
    T\kern-.1667em\lower.7ex\hbox{E}\kern-.125emX}}

    \makeatletter

\begin{document}
\flushbottom

\title{Communication-Aware Multi-Agent Reinforcement Learning for Cooperative UAV Deployment}

\author{
\IEEEauthorblockN{
Enguang Fan\textsuperscript{\dag},
Zihan Shan\textsuperscript{\dag},
Yifan Chen\textsuperscript{\dag},
Klara Nahrstedt\textsuperscript{\dag},
Matthew Caesar\textsuperscript{\dag},
Jae Kim\textsuperscript{\S}
}
\IEEEauthorblockA{
\textsuperscript{\dag}University of Illinois at Urbana-Champaign, \textsuperscript{\S}Boeing Research and Technology\\
Emails: \{enguang2, zshan2, yifanc3, klara, caesar\}@illinois.edu, \{jae.h.kim\}@boeing.com
}
}

\maketitle

\begin{abstract}
Autonomous unmanned aerial vehicle (UAV) swarms are increasingly used as rapidly deployable aerial relays and sensing platforms, yet practical deployments must operate under partial observability and intermittent peer-to-peer connectivity.
We present a graph-based multi-agent reinforcement learning framework trained under \emph{centralized training with decentralized execution} (CTDE): a centralized critic and global state are available only during training, while each UAV executes a shared policy using local observations and messages from nearby neighbors. {\color{black}Under restricted communication, neighbor relations are induced by an SNR-threshold connectivity graph.}
Our architecture encodes each agent's local state and nearby entities with an agent--entity attention module and aggregates inter-UAV messages with neighbor self-attention over a {\color{black}communication graph whose links are determined by a channel-based signal-quality threshold.}
We evaluate the framework on a cooperative relay-deployment task, DroneConnect. Experimental results show that the proposed method achieves an approximately $12\%$ relative increase in target coverage over MAPPO under restricted communication and partial observability, while remaining competitive with a mixed-integer linear programming (MILP)-based offline upper bound that assumes full node observability.
\end{abstract}


\section{Introduction}
\label{introduction}
Unmanned aerial vehicles (UAVs) are increasingly deployed as mobile sensing and communication platforms. Recent millimeter-wave (mmWave) sensing systems further demonstrate the potential of portable radar-based perception for drone-class platforms in contested environments and under poor visibility~\cite{mmPrism}.
A prominent application is to use autonomous UAVs as rapidly deployable \emph{aerial relays} when terrestrial infrastructure is damaged by natural disasters or overloaded during crowded events~\cite{yin2018uplink, Enguang_milcom}.
In scenarios such as wildfire monitoring and battlefield surveillance, UAVs may also operate beyond an operator's control range, requiring onboard autonomy and peer-to-peer coordination.
In such scenarios, UAV teams must decide where to position themselves to maximize sensing or communication coverage over dynamic areas of interest. These coverage and placement decisions can be formulated as optimization problems.
%

Many multi-UAV coverage and deployment tasks can be viewed through the lens of the Maximum Coverage Location Problem (MCLP), which is NP-hard and becomes computationally intractable in large or dynamic environments~\cite{sobouti2024uav}.
Moreover, real deployments are characterized by \emph{partial observability} (each UAV can sense only nearby entities) and \emph{communication constraints} {\color{black}(only UAV pairs whose link quality exceeds an SNR threshold can reliably exchange messages)}, making purely centralized controllers fragile and difficult to scale.

To address these challenges and optimize ground-node coverage by dynamically positioning UAVs, we develop a scalable multi-UAV deployment controller that integrates multi-agent reinforcement learning (MARL) with a graph-based environment representation.
We adopt \emph{centralized training with decentralized execution} (CTDE): during training, a centralized critic can access global information to stabilize learning, while during execution each UAV runs a shared policy using only local observations and peer-to-peer messages from communication neighbors.
Concretely, we model the environment as an \emph{agent--entity graph} and use attention-based embeddings to represent the state of each agent.
We evaluate our approach on a cooperative relay-deployment task (\emph{DroneConnect}) under full/partial observability and unrestricted/restricted communication. Our contributions are as follows.
\begin{itemize}[leftmargin=*]
    \item We propose a multi-agent reinforcement learning (MARL) framework for multi-UAV deployment under centralized training with decentralized execution (CTDE), partial observability, and {\color{black}SNR-limited communication constraints}.
    \item We introduce a dual-attention graph encoder that combines (i) agent--entity attention for local environment embedding and (ii) neighbor self-attention for inter-agent message aggregation.
    \item We demonstrate high-coverage decentralized relay deployment in the cooperative task (\emph{DroneConnect}), remaining competitive with an offline optimization-based upper bound and generalizing to unseen team sizes without fine-tuning.
\end{itemize}

The rest of this paper is structured as follows.
Section~\ref{sec:related_work} reviews prior research in learning and wireless communication for multi-UAV systems.
Section~\ref{sec:environment_modeling} introduces our environment embedding, message sharing, and CTDE learning design.
Section~\ref{sec:problem_modeling} presents the simulation scenarios.
We report evaluation results in Section~\ref{sec:performance} and conclude in Section~\ref{sec:conclusion}.

\begin{table*}[t]
\vspace*{0.07in}
\centering
\caption{Key notation used throughout the paper.}
\label{tab:notation}

\small
\renewcommand{\arraystretch}{1.2}

\begin{tabular}{@{} >{\centering\arraybackslash}m{0.48\textwidth}  m{0.48\textwidth} @{}}
\toprule
Symbol & Description \\
\midrule
$M$ / $N$ & number of UAVs / number of nodes (entities) \\
$\mathbf{p}_i(t)$ / $\mathbf{u}_j(t)$ & position of UAV $i$ / node $j$ at time $t$ \\
${r_s}$ / ${r_{cov}}$ & sensing radius / coverage radius \\ 
$\mathcal{N}_s(i)$ & entities sensed by UAV $i$ (within $r_s$) \\
$\mathcal{N}_c(i)$ & UAV neighbors of $i$ that satisfy the SNR threshold \\
$h_i$ / $m_i$ & latent embedding of UAV $i$ / aggregated message \\
FO/PO & full / partial observability \\
UC/RC & unrestricted / restricted communication \\
\bottomrule
\end{tabular}
\end{table*}

\section{Related Work}
\label{sec:related_work}
This section reviews prior work related to our study. We first summarize reinforcement-learning approaches for UAV control and then discuss communication mechanisms and graph-based representations for cooperative multi-agent coordination.

\textbf{Maximum Coverage Location Problem for Multiple UAVs}:
Early UAV-deployment research commonly frames the problem as a variant of the Maximum Coverage Location Problem (MCLP). When applied to UAV systems, MCLP captures the core challenge of selecting drone locations that maximize sensing or communication coverage under resource constraints. A representative formulation is the Maximum Coverage Facility Location Problem with Drones (MCFLPD) proposed by Chauhan et al.~\cite{CHAUHAN20191}, which models UAV deployment as a static mixed-integer program incorporating battery-limited range, energy consumption, and facility capacities. Despite its expressiveness, MCFLPD quickly becomes computationally expensive and therefore requires specialized heuristics to remain tractable.

\textbf{Reinforcement Learning for Multi-UAV Systems}: 
More broadly, reinforcement learning has been applied to wireless control under rapidly varying network conditions, including approaches that incorporate cross-layer information and adapt learned policies to network fluctuations~\cite{adaptiveVideoStreaming}. In UAV-assisted wireless networks, researchers have also demonstrated the utility of reinforcement learning algorithms~\cite{8494742, 8600371}. Lee et al.~\cite{Bella} introduced the DroneDR framework for UAV deployment within a centralized architecture; however, this design creates a single point of failure and is impractical in satellite-denied environments. Kaviani et al.~\cite{DeepCQ+} proposed DeepCQ+, a deep-reinforcement-learning-based routing protocol for highly dynamic mobile ad hoc networks. Our work addresses related scenarios but adopts a distributed approach. For wildfire monitoring, Julian et al.~\cite{wildfire-drone} employed deep Q-learning for UAV path planning.

\textbf{Multi-Agent Reinforcement Learning}: MARL encompasses fully cooperative, fully competitive, and mixed interaction settings. In this paper, we focus on decentralized control for a cooperative multi-UAV deployment task under partial observability and communication constraints.
Moreover, MARL approaches can be categorized as centralized, decentralized, or hybrid. Centralized approaches rely on a single controller for the entire multi-agent system, which becomes difficult to scale as the number of agents increases.
Decentralized approaches, such as Tampuu et al.'s Q-learning~\cite{marl-independent-cooperative}, employ independent Q-value functions for each agent but struggle in non-stationary environments. Centralized training with decentralized execution is another common approach, exemplified by algorithms such as COMA~\cite{coma}, BiCNet~\cite{bicnet}, and MADDPG~\cite{maddpg}, in which a centralized critic is available only during training.
{\color{black}Recent studies have also applied MARL techniques to wireless multi-UAV coordination~\cite{FedQMIX}. These works further highlight the importance of distributed coordination and graph-structured representations in wireless UAV systems. However, they often target different wireless objectives or learning settings, whereas our work focuses on CTDE-based decentralized relay deployment under partial observability and communication constraints and is designed to generalize across varying agent counts.}

\textbf{Communication Mechanisms Between Agents}: Wireless coordination can be improved at multiple layers. At the physical layer, channel-aware optimization techniques such as rate splitting and beamforming improve multi-user and multicast transmission efficiency under fading~\cite{rateSplitting}. At the agent layer, many MARL approaches do not model explicit inter-agent communication. Differentiable communication protocols, such as CommNet~\cite{commnet} and VAIN~\cite{vain}, improve coordination through attention-based message passing. We likewise use scaled dot-product attention for inter-agent communication.
Real-world scenarios often impose communication limits based on proximity, as seen in TarMAC~\cite{tarmac}. DGN~\cite{dgn} allows agents to communicate with their nearest neighbors, aligning with practical drone-swarm operations. {\color{black}However, these methods are primarily designed for general multi-agent settings rather than UAV-specific communication networks governed by radio-frequency wireless channels.}


\textbf{Graph Neural Networks}: Graphs naturally model multi-agent systems, with nodes representing agents. GNNs, such as message-passing neural networks~\cite{mpnn} and the Graph Attention Network (GAT)~\cite{graph-attention-net}, employ trainable weights to propagate features among nodes. Agarwal et al.~\cite{transferrable-cooperative-marl} introduced entity graphs to represent the environment in fully cooperative settings. OpenAI studied the emergence of complex behavior through multi-agent reinforcement learning~\cite{hide-and-seek}. In our work, we adopt an agent--entity graph to aggregate environmental information for decentralized cooperative UAV deployment.

\begin{figure*}[t]
    \centering
    \includegraphics[width=\textwidth]{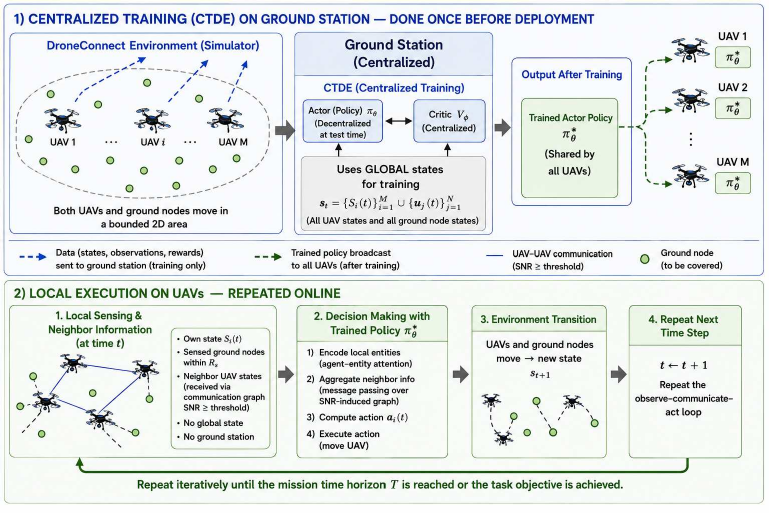}
    \caption{Overview of the proposed communication-aware CTDE framework.}
    \label{fig:system_description}
\end{figure*}

\section{Communication-Aware CTDE Framework}
\label{sec:environment_modeling}
Figure~\ref{fig:system_description} summarizes the proposed communication-aware CTDE framework. During centralized training, a critic uses global state to guide learning of a shared actor. The trained actor is then deployed to all UAVs for decentralized execution using only local observations and peer-to-peer messages; under restricted communication, these messages come only from neighbors whose links satisfy the SNR threshold.

To model large multi-agent environments efficiently, we represent the swarm and its surroundings as an \emph{agent--entity graph} $G=(V,E)$, where vertices correspond to UAV agents and observable environment entities (e.g., ground nodes), and edges encode sensing and communication relationships.
Each UAV $i$ forms (i) a sensed-entity set $\mathcal{N}_s(i)$ consisting of entities within its sensing radius and (ii) a communication-neighbor set $\mathcal{N}_c(i)$ consisting of UAVs whose links satisfy the {\color{black}SNR-threshold connectivity rule}.
This separation allows us to model \emph{partial observability} (via $\mathcal{N}_s$) and \emph{restricted communication} {\color{black}(via the SNR-induced set $\mathcal{N}_c$)} in a unified way.


\subsection{Message Passing Over the Communication Graph} \label{channel_model}
We use message passing to aggregate information among UAVs over the communication graph.
Let $h_i^{(k)}$ denote UAV $i$'s latent embedding after $k$ message-passing rounds, with $h_i^{(0)}$ initialized from its local observation embedding (Section~\ref{Environment Embedding}).
A generic message-passing round can be written as
\begin{align}
    m^{(k)}_i &= f_{\text{agg}}^{(k)}\!\left(h^{(k-1)}_i,\;\{h^{(k-1)}_j : j \in \mathcal{N}_c(i)\}\right), \quad 1\leq k \leq K \label{eq:mp1} \\
    h^{(k)}_i &= f_{\text{upd}}\!\left(h^{(k-1)}_i,\; m^{(k)}_i\right), \label{eq:mp2}
\end{align}
where $m^{(k)}_i$ is the aggregated message and $K$ is the number of message-passing rounds. Here, $f_{\text{agg}}^{(k)}(\cdot)$ denotes a permutation-invariant aggregation operator over neighbor embeddings (instantiated as attention-weighted aggregation in Section~\ref{Inter-Agent Message Sharing}), and $f_{\text{upd}}(\cdot)$ is a learnable update function (e.g., a multilayer perceptron or gated recurrent unit) that fuses the previous embedding with the aggregated message.
In unrestricted communication (UC), $\mathcal{N}_c(i)$ contains all other UAVs; in restricted communication (RC), {\color{black}the communication graph is rebuilt from the pairwise SNRs according to the empirical UAV-to-UAV mmWave channel model of Polese et al.~\cite{polese2020experimental}:
\begin{align}
    \mathrm{SNR}_{ij}(t) &= \frac{P_t\, G_t G_r\, g_{ij}(t)}{P_n}, \label{eq:snr_polese} \\
    g_{ij}(t) &= 10^{-PL_{ij}/10}, \label{eq:gain_polese} \\
    PL_{ij}(t) &= PL_{\mathrm{FS},d_0}(f_c)
    + 10 n_{\mathrm{CI}} \log_{10}\!\left(\frac{d_{ij}(t)}{d_0}\right)
    + X_{\sigma,ij}, \label{eq:pl_polese} \\
    PL_{\mathrm{FS},d_0}(f_c) &= 20 \log_{10}\!\left(\frac{4\pi f_cd_0}{c}\right), \label{eq:fspl_polese} \\
    X_{\sigma,ij} &\sim \mathcal{N}(0,\sigma^2). \label{eq:shadow_polese}
\end{align}

An edge is present if and only if
\[
    A_{ij}(t)=\mathbb{I}[\mathrm{SNR}_{ij}(t)\ge \tau],
    \qquad
    \mathcal{N}_c(i)=\{j\neq i: A_{ij}(t)=1\}.
\]

$\mathrm{SNR}_{ij}(t)$ denotes the signal-to-noise ratio from UAV $i$ to UAV $j$ at time $t$; $P_t$ is the transmit power; $G_t$ and $G_r$ are the transmit and receive antenna gains, respectively; $g_{ij}(t)$ is the linear-scale channel gain; $PL_{ij}(t)$ is the large-scale path loss in dB; $PL_{\mathrm{FS},d_0}(f_c)$ is the free-space path loss at the reference distance $d_0 = 1\,\mathrm{m}$ for carrier frequency $f_c$; $d_{ij}(t)$ is the physical distance between UAVs $i$ and $j$ at time $t$, given by $\|\mathbf{p}_i(t)-\mathbf{p}_j(t)\|_2$; $n_{\mathrm{CI}}$ is the close-in path-loss exponent; $X_{\sigma,ij}$ is the shadowing term in dB; $\sigma$ is the standard deviation of the shadowing term; $c$ is the speed of light; and $P_n$ is the receiver noise power over the system bandwidth.

}

\subsection{Environment Embedding}
\label{Environment Embedding}
Each UAV $i$ maintains a local state $S_i$ (e.g., position and velocity). We assume that this kinematic state is provided by the platform’s localization stack; complementary wireless and inertial localization systems can obtain position estimates without requiring a pre-mapped infrastructure layout~\cite{sie2025crowdsourcingubiquitousindoorlocalization}. {\color{black}In the DroneConnect implementation, $S_i(t)=[p_i^x(t),p_i^y(t),v_i^x(t),v_i^y(t)]$, while each observed node $l$ is encoded by a relative feature vector $\mathbf{x}_{i,l}(t)=[u_l^x-p_i^x,\;u_l^y-p_i^y,\;v_l^x,\;v_l^y,\;\|\mathbf{u}_l-\mathbf{p}_i\|_2]$. The actor therefore receives the local kinematic state together with padded node features and the current communication mask induced by the SNR graph.}
We encode the UAV state and entity features via
\begin{align}
    \mathbf{h}^{a}_i &= f_a(S_i), \\
    \mathbf{e}_{i,l} &= f_e(\mathbf{x}_{i,l}), \quad l \in \mathcal{N}_s(i),
\end{align}
where $\mathbf{x}_{i,l}$ denotes the feature vector of entity $l$ as observed by UAV $i$. {\color{black}We instantiate both the agent-state encoder and the entity encoder as one-hidden-layer multilayer perceptrons with ReLU activations. Specifically, $f_a$ encodes the local UAV state $S_i(t)=[p_i^x,p_i^y,v_i^x,v_i^y]$, while $f_e$ encodes each observed node feature $\mathbf{x}_{i,l}(t)=[\Delta x_{i,l},\Delta y_{i,l},v_l^x,v_l^y,\|\mathbf{u}_l-\mathbf{p}_i\|_2]$.}

To obtain a fixed-size environment summary that is invariant to the number of sensed entities, we apply scaled dot-product attention with the UAV embedding as the query and entity embeddings as keys/values:
{\setlength{\jot}{8pt} 
\begin{align}
    \mathbf{q}_i &= \mathbf{W}_q \mathbf{h}^{a}_i, \\
    \mathbf{k}_{i,l} &= \mathbf{W}_k \mathbf{e}_{i,l}, \quad l \in \mathcal{N}_s(i), \\
    \mathbf{v}_{i,l} &= \mathbf{W}_v \mathbf{e}_{i,l}, \quad l \in \mathcal{N}_s(i), \\
    \alpha_{i,l} &= \frac{\exp\!\left(\mathbf{q}_i^\top \mathbf{k}_{i,l}/\sqrt{d_k}\right)}
    {\sum_{l' \in \mathcal{N}_s(i)} \exp\!\left(\mathbf{q}_i^\top \mathbf{k}_{i,l'}/\sqrt{d_k}\right)}, \\
    \mathbf{E}^{agg}_i &= \sum_{l \in \mathcal{N}_s(i)} \alpha_{i,l}\, \mathbf{v}_{i,l}, \\
    h^{(0)}_i &= \left[\mathbf{h}^{a}_i;\mathbf{E}^{agg}_i\right].
\end{align}
}

Here, $d_k$ is the key dimension and $[;]$ denotes concatenation.
The resulting $h^{(0)}_i$ is used as the per-agent input to the policy and as the initialization for communication message passing.

\subsection{Inter-Agent Message Sharing}
\label{Inter-Agent Message Sharing}
UAV $i$ aggregates messages from its communication neighbors $\mathcal{N}_c(i)$ using self-attention.
Let $\tilde{\mathcal{N}}_c(i)=\mathcal{N}_c(i)\cup\{i\}$ denote the neighbor set with a self-loop.
For round $k$, we compute
{\setlength{\jot}{8pt} 
\begin{align}
    \mathbf{q}^{c}_i &= \mathbf{W}^{c}_q h^{(k-1)}_i, \\
    \mathbf{k}^{c}_j &= \mathbf{W}^{c}_k h^{(k-1)}_j, \quad j \in \tilde{\mathcal{N}}_c(i), \\
    \mathbf{v}^{c}_j &= \mathbf{W}^{c}_v h^{(k-1)}_j, \quad j \in \tilde{\mathcal{N}}_c(i), \\
    \beta_{i,j} &= \frac{\exp\!\left((\mathbf{q}^{c}_i)^\top \mathbf{k}^{c}_j/\sqrt{d_k}\right)}
    {\sum_{j' \in \tilde{\mathcal{N}}_c(i)} \exp\!\left((\mathbf{q}^{c}_i)^\top \mathbf{k}^{c}_{j'}/\sqrt{d_k}\right)}, \\
    m^{(k)}_i &= \sum_{j \in \tilde{\mathcal{N}}_c(i)} \beta_{i,j}\, \mathbf{v}^{c}_j, \\
    h^{(k)}_i &= f_{\text{upd}}\!\left(h^{(k-1)}_i,\; m^{(k)}_i\right)
\end{align}
}
During decentralized execution, each UAV performs this aggregation using only messages received from $\mathcal{N}_c(i)$, matching the RC setting. {\color{black}Equivalently, UAV $i$ broadcasts its latent embedding only along links whose current SNR exceeds $\tau$ and then applies self-attention over the remaining neighbors plus its own self-loop.}

\subsection{Centralized Training with Decentralized Execution (CTDE)}
\label{subsec:ctde}
We train the swarm under CTDE.
During training, we additionally use a centralized critic that has access to global information, e.g.,
\begin{equation}
    V_{\phi}\!\left(\mathbf{s}(t)\right), \qquad \mathbf{s}(t)=\{S_i(t)\}_{i=1}^{M} \cup \{\mathbf{u}_j(t)\}_{j=1}^{N}.
\end{equation}
The critic is used only for learning; at test time, UAVs execute $\pi_\theta$ without access to $\mathbf{s}(t)$ or any centralized coordinator.
During execution, each UAV $i$ samples an action from the decentralized actor $\pi_{\theta}$:
\begin{equation}
    a_i(t) \sim \pi_{\theta}\!\left(\cdot \mid o_i(t), h^{(K)}_i(t)\right),
\end{equation}
where $o_i(t)$ is the local observation and $h^{(K)}_i(t)$ is the final embedding after $K$ communication rounds.

\section{Scenarios and Tasks}
\label{sec:problem_modeling}
We study multi-UAV deployment under partial observability and {\color{black}SNR-limited communication}.
Our focus is a cooperative relay deployment task (\emph{DroneConnect}).
We also compare against a static optimization-based formulation as a reference upper bound.

\subsection{Optimization-Based Static View}
\label{subsec:optimization-based static view}
A common abstraction of coverage and relay placement maximizes the demand covered within a service radius while penalizing relocation costs. Facility locations represent UAV positions, and demand points represent ground nodes.
Let $\mathbf{p}_i^0$ be the current position of UAV $i$, and let $\mathbf{p}_i$ be its placement decision.
A simplified maximum-coverage objective can be written as
\begin{align}
\max_{\{\mathbf{p}_i\}_{i=1}^{M}}\;\; & \sum_{j=1}^{N} w_j\, z_j \;-\; \alpha \sum_{i=1}^{M} \left\|\mathbf{p}_i-\mathbf{p}_i^0\right\|_2 \label{eq:mclp_obj}\\
\text{s.t.}\;\; & z_j = \mathbb{I}\!\left[\min_{i \in \{1,\dots,M\}} \left\|\mathbf{u}_j-\mathbf{p}_i\right\|_2 \le r_{cov}\right], \;\; j=1,\dots,N \nonumber
\end{align}
where $w_j$ is a node priority weight and $r_{cov}$ is the service/coverage radius.
This problem is NP-hard; to provide an optimization-based reference, we discretize candidate UAV locations and solve a mixed-integer linear programming (MILP) formulation of (\ref{eq:mclp_obj}) offline, which serves as an approximate upper bound on the attainable coverage for static snapshots. However, MILP is computationally intensive and is therefore not suitable for online control.

\subsection{DroneConnect Scenario}
\label{Cooperative Scenario}
\begin{figure}
\begin{center}
  \frame{\includegraphics[width=.6\linewidth]{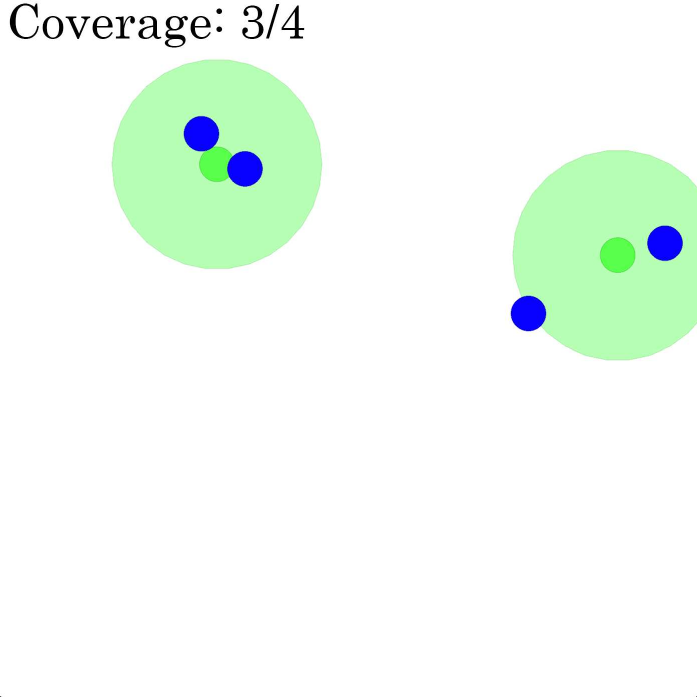}}
\end{center}
  \caption{DroneConnect environment with 2 UAV and 4 mobile ground nodes.}
\label{fig:drone-connect-render}
\vspace{-2em}
\end{figure}

DroneConnect models a team of UAV relays that reposition to provide coverage for mobile ground nodes (Fig.~\ref{fig:drone-connect-render}).

\subsubsection{Action Space}
For UAV $i$, the continuous action is a 2D force (or acceleration) command, $a_i=(F_x,F_y)$, that updates its velocity and position.

\subsubsection{Observability and Communication Settings}
We evaluate four settings that combine observation and communication constraints:
\begin{itemize}[leftmargin=*]
    \item \textbf{FO (full observability)}: each UAV observes all ground-node states (and UAV states).
    \item \textbf{PO (partial observability)}: each UAV observes only entities within sensing radius $r_s$ (i.e., $\mathcal{N}_s(i)$) through onboard sensing.
    \item \textbf{UC (unrestricted communication)}: all UAVs can exchange messages over a complete communication graph.
    \item \textbf{RC (restricted communication)}: UAVs communicate only when their pairwise link satisfies {\color{black}$\mathrm{SNR}_{ij}(t)\ge \tau$}; {\color{black}equivalently, the RC graph is defined by $A_{ij}(t)=\mathbb{I}[\mathrm{SNR}_{ij}(t)\ge \tau]$}.
\end{itemize}

{\color{black}For the restricted-communication experiments, we use the empirical UAV-to-UAV channel model described in Section~\ref{channel_model}. In the experimental environment, we set the carrier frequency to $f_c=60.48\,\mathrm{GHz}$, the close-in path-loss exponent to $n_{\mathrm{CI}}=2.25$, the shadowing standard deviation to $\sigma=3.56\,\mathrm{dB}$, the connectivity threshold to $\tau=10\,\mathrm{dB}$, the transmit power to $P_t=20\,\mathrm{dBm}$, and the antenna gains to $G_t=G_r=10\,\mathrm{dBi}$. This parameterization yields a sparse connectivity pattern in the $100\,\mathrm{m}\times 100\,\mathrm{m}$ workspace, approximating a realistic UAV-swarm networking regime.}

\subsubsection{Reward}
The DroneConnect task is fully cooperative: all UAVs share a common team reward during centralized training and execute decentralized policies at test time.
Let $d_j(t)=\min_i \|\mathbf{u}_j(t)-\mathbf{p}_i(t)\|_2$ be the distance from node $j$ to its nearest UAV, and let $c_j(t)=\mathbb{I}[d_j(t)\le r_{cov}]$ indicate whether node $j$ is covered.
We use the normalized reward
\begin{equation}
\label{eq:1_reward}
r(t)=\lambda_{cov}\cdot \frac{1}{N}\sum_{j=1}^{N} c_j(t)\;-\;\lambda_{dist}\cdot \frac{1}{N}\sum_{j=1}^{N} \frac{d_j(t)}{r_{cov}},
\end{equation}
where $\lambda_{cov}$ and $\lambda_{dist}$ trade off coverage quantity and service quality.

\section{Evaluation}
\label{sec:performance}
We evaluate the proposed graph-based CTDE MARL framework on the DroneConnect task introduced in Section~\ref{sec:problem_modeling}.

\subsection{Experimental Setup}
We report representative settings here to support reproducibility.
In DroneConnect, UAVs and nodes move in a bounded 2D area of size $100\times 100$ with time step $\Delta t=0.1\,\mathrm{s}$ and episode length $T=200$ steps.
Each UAV has a sensing radius $r_s$. {\color{black}In RC, UAVs communicate according to the SNR-threshold graph $A_{ij}=\mathbb{I}[\mathrm{SNR}_{ij}\ge \tau]$; in UC, all-to-all messaging is allowed}.
We train a proximal policy optimization (PPO)-style CTDE actor--critic with a centralized critic and decentralized actors for $2\times 10^6$ environment steps and evaluate it over 50 episodes.
Unless otherwise stated, we report averages over 3 random seeds.
For communication, each UAV transmits a $d{=}64$-dimensional message embedding to each neighbor per time step; thus the per-step communication cost scales with the average degree of the RC graph.

\subsection{DroneConnect Results}
Here, $M$ denotes the number of UAVs and $N$ denotes the number of ground nodes to be covered.
The evaluation metric is the \emph{average coverage ratio}: the fraction of ground nodes covered at each time step, averaged over an episode.

\begin{figure}[t]
\centering
\vspace*{0.04in}
\includegraphics[width=\columnwidth]{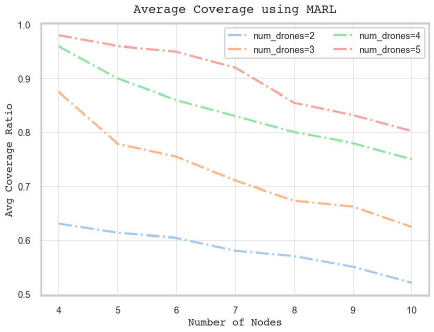}
\caption{Average coverage per time step as the numbers of UAVs ($M$) and nodes ($N$) vary (FO+UC setting).}
\label{fig:marl_coverage}
\end{figure}


\subsubsection{Coverage Results and Ablations}
Table~\ref{tab:relay_result} summarizes coverage under different observability and communication constraints.
Overall, our CTDE approach maintains strong coverage under partial observability and restricted communication. {\color{black}The multi-agent proximal policy optimization (MAPPO) adaptation achieves comparable coverage under full observability (FO), but its coverage drops by approximately $10\%$ under partial observability (PO), consistent with the increased difficulty of decentralized control under limited local information. Meanwhile, our CTDE approach remains competitive with the static MILP upper bound and the centralized single-agent RL baseline, both of which assume full observability but are tailored to fixed team sizes and are less scalable than our method.}
We also report two minimal ablations: disabling inter-UAV communication and replacing entity attention with mean pooling; both degrade performance in the challenging RC+PO (restricted communication and partial observability) setting. Figure~\ref{fig:marl_coverage} additionally shows how coverage varies with the numbers of UAVs \(M\) and ground nodes \(N\) under FO+UC. Coverage generally decreases as \(N\) increases for a fixed \(M\), while larger UAV teams maintain higher coverage.


\begin{table}[t]
\vspace*{0.04in}
\centering
\footnotesize
\setlength{\tabcolsep}{2pt}
\renewcommand{\arraystretch}{1.05}
\begin{tabular}{@{}lccccc@{}}
\toprule
Method & $M$ & $N$ & Comm & Obs & Coverage \\
\midrule
Ours & 3 & 6 & UC & FO & $0.76 \pm 0.01$ \\
Ours & 3 & 6 & RC & FO & $0.74 \pm 0.02$ \\
Ours & 3 & 6 & UC & PO & $0.72 \pm 0.02$ \\
Ours & 3 & 6 & RC & PO & $0.71 \pm 0.02$ \\
Ours & 5 & 10 & UC & FO & $0.79 \pm 0.02$ \\
Ours & 5 & 10 & RC & FO & $0.77 \pm 0.02$ \\
Ours & 5 & 10 & UC & PO & $0.78 \pm 0.02$ \\
Ours & 5 & 10 & RC & PO & $0.76 \pm 0.02$ \\
\midrule
No comm. & 5 & 10 & RC & PO & $0.65 \pm 0.03$ \\
No entity attn. & 5 & 10 & RC & PO & $0.63 \pm 0.03$ \\
\midrule
Static MILP ref. & 3 & 6 & -- & -- & $0.81$ \\
Static MILP ref. & 5 & 10 & -- & -- & $0.84$ \\
Centralized RL & 3 & 6 & -- & FO & $0.75 \pm 0.02$ \\
Centralized RL & 5 & 10 & -- & FO & $0.79 \pm 0.02$ \\
MAPPO & 5 & 10 & RC & FO & $0.77 \pm 0.03$ \\
MAPPO & 5 & 10 & RC & PO & $0.68 \pm 0.02$ \\
\bottomrule
\end{tabular}
\caption{DroneConnect ground-node coverage ratio (mean $\pm$ std over 3 seeds). UC/RC: unrestricted/restricted communication; FO/PO: full/partial observability.}
\label{tab:relay_result}
\vspace{0.02in}
\end{table}
\subsubsection{Qualitative Coordination and Overlap}
\begin{figure}[!htb]
    \centering
    \begin{subfigure}[b]{0.49\linewidth}
        \centering
        \includegraphics[width=\linewidth]{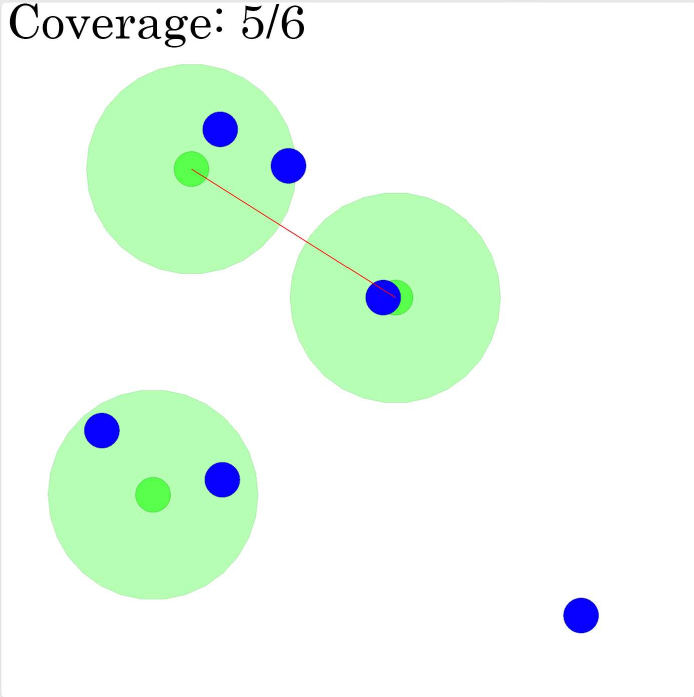}
        \caption{$M=3, N=6$}
    \end{subfigure}
    \hfill
    \begin{subfigure}[b]{0.49\linewidth}
        \centering
        \includegraphics[width=\linewidth]{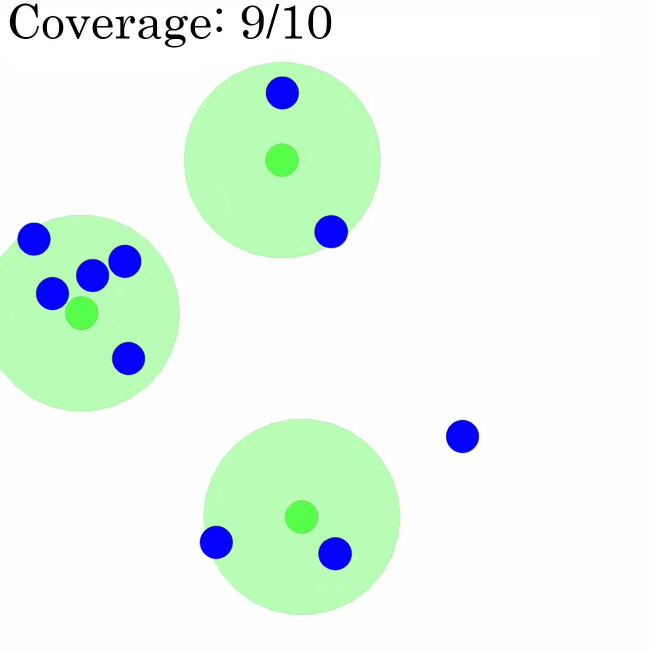}
        \caption{$M=3, N=10$}
    \end{subfigure}

    \vspace{0.5em}

    \begin{subfigure}[b]{0.49\linewidth}
        \centering
        \includegraphics[width=\linewidth]{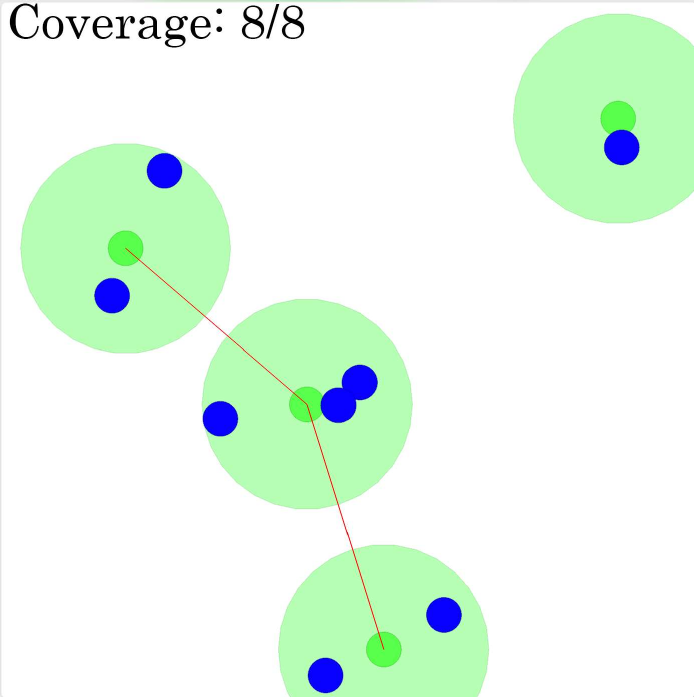}
        \caption{$M=4, N=8$}
    \end{subfigure}
    \hfill
    \begin{subfigure}[b]{0.49\linewidth}
        \centering
        \includegraphics[width=\linewidth]{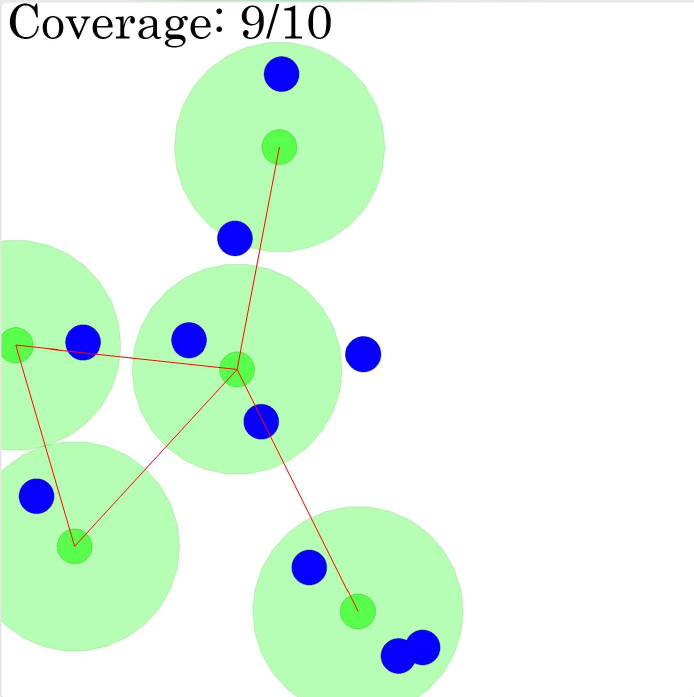}
        \caption{$M=5, N=10$}
    \end{subfigure}
    \caption{Representative DroneConnect snapshots showing division of RC coverage tasks.}
    \label{fig:drone_relay_emergent}
\end{figure}

In addition to the coverage ratio, we quantify coordination by measuring the \emph{coverage overlap rate}, defined as the fraction of covered nodes that are simultaneously within $r_{cov}$ of more than one UAV.
In the challenging RC+PO setting with $M{=}5,N{=}10$, our learned policy achieves a low overlap rate of $0.12$, compared with $0.19$ for the no-communication ablation, indicating better division of coverage responsibilities.
Figure~\ref{fig:drone_relay_emergent} illustrates representative trajectories and communication links under different team sizes.

\section{Conclusion}
\label{sec:conclusion}
We presented a \emph{centralized training with decentralized execution} (CTDE) multi-agent reinforcement learning framework for cooperative UAV deployment under partial observability and communication constraints.
Our method represents the environment as an agent--entity graph and uses dual attention: agent--entity attention for local environment embedding and neighbor self-attention for inter-UAV message aggregation.
During execution, each UAV runs a decentralized policy using only local observations and peer-to-peer messages, without any centralized coordinator.

In the cooperative DroneConnect task, our approach achieves high coverage under restricted communication and partial observability while remaining competitive with the static MILP upper bound.
We also showed that the learned policy can generalize zero-shot to different team sizes in DroneConnect.
Future work will consider more detailed wireless quality of service (QoS) models with fading and interference, stronger MARL baselines, and explicit analysis of communication-cost constraints.

\bibliographystyle{ieeetr}
\bibliography{Main_Bib}

\end{document}